\documentclass[reprint,
 amsmath,amssymb,
 aps,
]{revtex4-1}

\usepackage{graphicx}
\usepackage{dcolumn}
\usepackage{bm}

\usepackage[usenames, dvipsnames]{color}
\usepackage{soul}
\usepackage{float}
\usepackage{siunitx}

\begin{document}
\title{Transmitting more than 10 bit with a single photon}

\author{T. B. H. Tentrup$^1$}
\author{T. Hummel$^{1}$}
\author{T. A. W. Wolterink$^{1,2}$}
\author{R. Uppu$^1$}
\author{A. P. Mosk$^{1}$}
\author{P. W. H. Pinkse$^1$}

\affiliation{$^1$ Complex Photonic Systems (COPS), MESA+ Institute for Nanotechnology, University of Twente, P.O. Box 217, 7500 AE Enschede, The Netherlands}
\affiliation{$^2$ Laser Physics and Nonlinear Optics (LPNO), MESA+ Institute for Nanotechnology, University of Twente, P.O. Box 217, 7500 AE Enschede, The Netherlands}

\date{\today}

\begin{abstract}
Encoding information in the position of single photons has no known limits, given infinite resources. Using a heralded single-photon source and a Spatial Light Modulator (SLM), we steer single photons to specific positions in a virtual grid on a large-area spatially resolving photon-counting detector (ICCD). We experimentally demonstrate selective addressing any location (symbol) in a 9072 size grid (alphabet) to achieve 10.5 bit of mutual information between the sender and receiver per detected photon. Our results set the stage for very-high-dimensional quantum information processing.
\end{abstract}

\maketitle

\section{Introduction}

Its weak interaction with the environment makes light ideal for sharing information between distant parties. For this reason, light is used to transmit information all around the world. With the advent of single-photon sources, a new class of applications has emerged. Due to their quantum properties, single photons are used to entangle quantum systems or to do quantum cryptography \cite{Nielsen:2011:QCQ:1972505}. One famous example is Quantum Key Distribution (QKD) using the BB84 protocol \cite{bb84} to securely build up a secret shared key between Alice and Bob. The security of this method is based on the no-cloning theorem \cite{Noclone}, which forbids copying quantum states. The standard implementation of the BB84 protocol uses the two-dimensional polarization basis to encode information in photons. Therefore the alphabet contains only two symbols "0" and "1" limiting the information content per photon to $1$ bit. Increasing the dimension of the basis using a large alphabet increases the information content per photon together with an improvement in the security \cite{RevModPhys.81.1301,PhysRevLett.88.127902,PhysRevA.61.062308}. This is the motivation to employ larger alphabets using orbital angular momentum \cite{1367-2630-17-3-033033,PhysRevA.88.032305}, time binning \cite{PhysRevLett.98.060503,1367-2630-17-2-022002} or spatial translation \cite{PhysRevLett.96.090501,PhysRevA.77.062323,Dixon2012}. It has recently been shown that, assuming a sender-receiver configuration with apertures of finite size, a diffraction-limited spot translated in space has a higher capacity limit than orbital-angular-momentum states \cite{Zhao2015}. This makes spatial encoding of light the preferred candidate for increasing the information content per photon. Interestingly, given infinite resources, there is no known upper bound for the information content transmitted by single photons. For example, using one mole $\left(6.022 \cdot 10^{23} \right)$ of ideal position-sensitive single-photon detectors leads to an information content of $79$ bit per detected photon. Clearly, this is out of reach in a practical situation. A very relevant question therefore is what can be realised experimentally.
\newline
\par
In this article we report our experiment in which we have encoded more than $10$ bit of information into a single photon. Using spatial encoding, we have employed $2^3$ times more symbols than previous work, which reported $7$ bit per photon as highest value \cite{Dixon2012}.

\section{Result}
\label{Result}

\begin{figure}[!htbp]
\centering
\includegraphics[width=8cm]{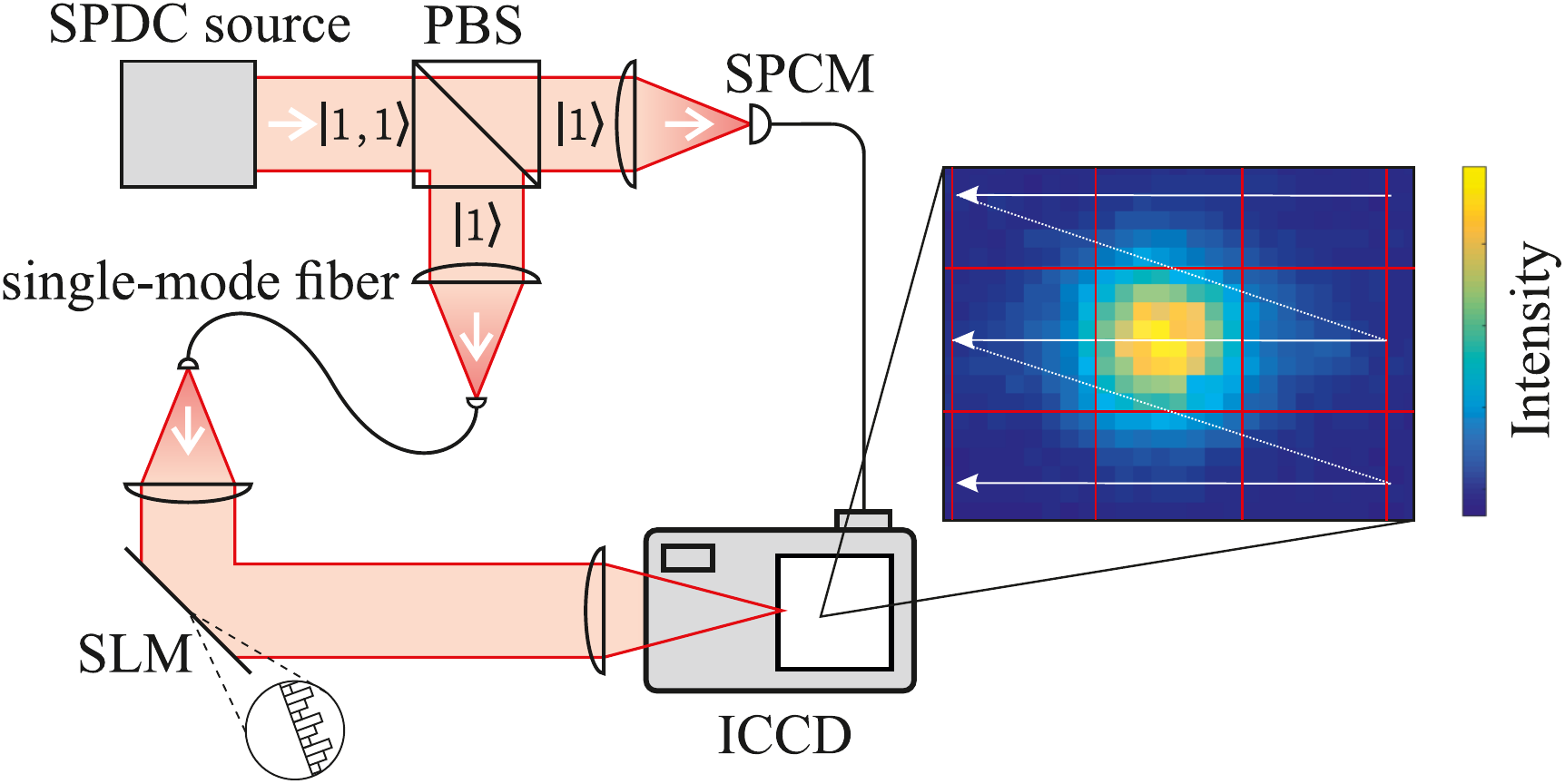}
\caption{Schematic representation of the setup. The Type-II Spontaneous Parametric Down-Conversion (SPDC) source produces photon pairs, which are split by a Polarising Beam Splitter (PBS). One of the photons is detected by a Single-Photon Counting Module (SPCM) acting as a herald for an Intensified CCD (ICCD). The other photon is fiber-coupled and is incident on a Spatial Light Modulator (SLM). The Fourier image is projected on the ICCD. The position of the focus is scanned by the blazed grating on the SLM, indicated by the arrows. An accumulated focus is shown in a zoom-in of the ICCD image integrated over an average of $1000$ photons. The red lines show the $8\times8$ pixel binning of the symbols.}
\label{fig:setup}
\end{figure}

The spatial encoding of information is realized on a rectangular, virtual grid formed by binning the pixels of a camera. The experimental setup is schematized in Fig. \ref{fig:setup} and details are given in Methods. We used a grid of $112\times81$ areas of binned pixels, which are the $9072$ symbols of our alphabet. This value corresponds to a maximum information content of $\log_2(9072)=13.15$ bit.

\begin{figure}[hbt]
\centering
\includegraphics[width=8cm]{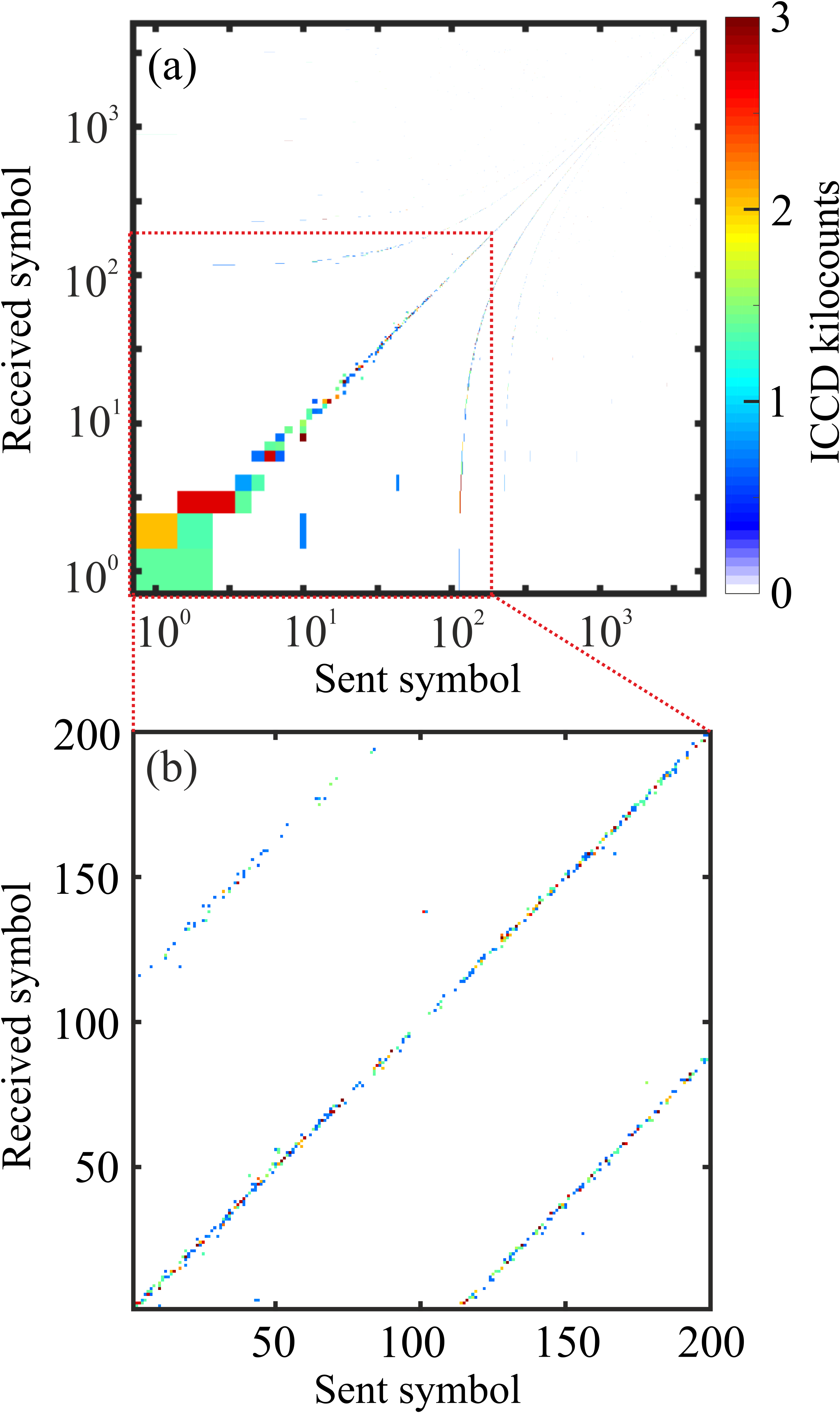}
\caption{Measured ICCD counts in each of the measured symbols as a function of the sent symbol with a binning size of $8\times 8$ pixels. The exposure time was \SI{0.6}{\second} for each symbol. Graph (a) illustrates the correlation between all $9072$ symbols in a log-log plot. The graph (b) shows the measured correlation between the first $200$ symbols. The measurement samples the joint probability distribution $P(X,Y)$.} 
\label{fig:result}
\end{figure}

The light is directed to distinct symbols on the grid by scanning the focus using a SLM as a blazed grating. We assume that the sender has an alphabet $X$ and the receiver an alphabet $Y$. To characterize the system, we sampled the joint probability distribution $P(X,Y)$, which indicates the probabilities $p(x,y)$ to detect a certain symbol $y$ out of the alphabet $Y$ if a symbol $x$ out of the alphabet $X$ was sent. The result is plotted in Fig. \ref{fig:result}. The readout noise of the ICCD was reduced by applying a threshold on the measured signal to only show the intensified signal. A diagonal line in the plot of the sent symbol versus the received would indicate maximal correlation and thus $13.15$ bit of information. Indeed there is a strong correlation between the sent and received symbol set visible in graph (a), which shows the hole alphabet in a log-log plot. Graph (b) depicts a zoom into the first $200$ of the $9072$ symbols. Due to crosstalk between the symbols, off-diagonal lines are visible, which correspond to photons hitting the symbol above or below the target. The distance of $112$ symbols between these lines and the diagonal corresponds to the column length of the grid written on the camera. The signal to the left and right of the diagonal is caused by the crosstalk between the left and right symbol on the grid. Noise can also arise from dark counts of the ICCD.

To quantify the information content per photon, we calculate the mutual information between sender and receiver. The mutual information $I(X:Y)$ is the measure of the average reduction in uncertainty about a sent symbol set $X$ that results from learning the value of the received symbol set $Y$; or vice versa, the average amount of information that $X$ conveys about $Y$ \cite{mackay2003information}. The mutual information per detection event of the sender-receiver system is mathematically represented as
\begin{equation}
I(X:Y)=\sum_{x \in X,y\in Y} p(x,y) \log_2 \left(\frac{p(x,y)}{p(x)p(y)} \right),
\label{eq:mutualinformation}
\end{equation}  
where $p(x,y)$ is the probability that symbol $y$ is received when symbol $x$ is send and $p(x)$ and $p(y)$ are the overall probability to measure symbol $x$ or symbol $y$. Theoretically, the mutual information depends on the number of symbols $N$, which has a maximum of $I_\text{max}=\log_2(N)$. The maximum number of symbols is limited by the finite size of the CCD. Using the presented binning size of the detection areas of $8\times8$, our theoretical limit is $13.15$ bit.

\begin{figure}[htbp]
\centering
\includegraphics[width=8cm]{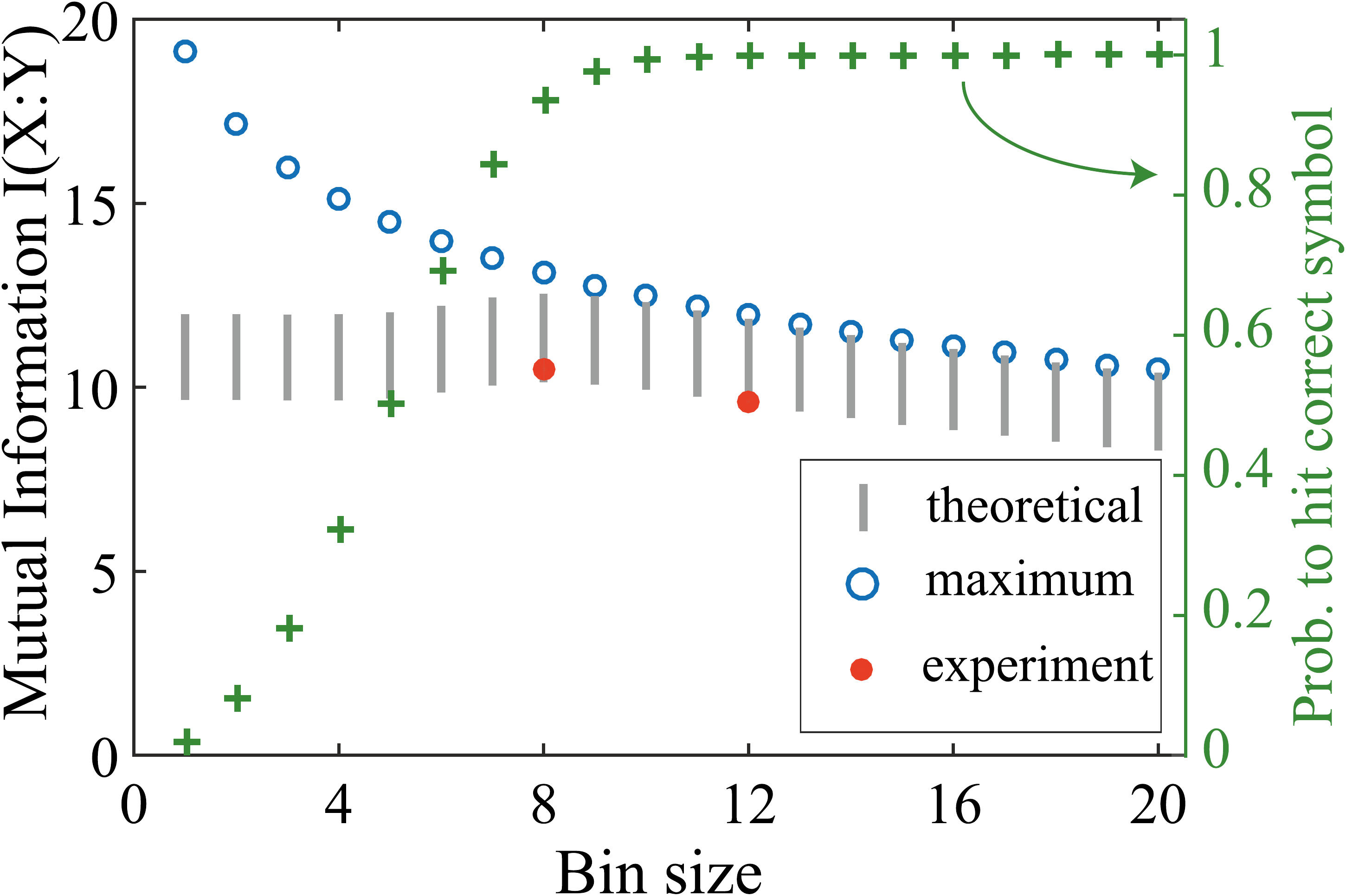}
\caption{Dependence of the mutual information and the average hit probability in the correct symbol on the binning size of the detection areas. The blue circles represent the theoretical limit $I_\text{max}$ given no noise or crosstalk. The red dots correspond to the measured mutual information for $8 \times 8$ and $12\times 12$ pixel bin size. The theoretical mutual information is shown as gray bars in the figure, corrected for a signal-to-dark-count photon ratio between 10 and 100. The green $+$ markers illustrate the average hit probability in the correct area for a finite focal diameter with FWHM of 8 pixels as shown in Fig. \ref{fig:setup}.}
\label{fig:MIvsP}
\end{figure}

In reality, the mutual information is not only limited by the number of symbols, but also by the crosstalk between the symbols, arising from diffraction-limited focal spots, as well as the detector and background noise. In order to reduce the crosstalk between the symbols, the binning size of the detection areas can be increased. However, the limited number of pixels on the detector reduces the number of symbols. The blue circles in Fig. \ref{fig:MIvsP} show the dependence of the maximum mutual information for symbols made with different pixel bin sizes. The measured mutual information for $8 \times 8$ and $12\times 12$ pixel bin sizes are represented as red circles, which are lower than the theoretical limit. The green $+$ markers in Fig.~\ref{fig:MIvsP} depict the average hit probability in our experiment. By accounting for this crosstalk together with finite signal-to-dark-photocount ratio between $10$ and $100$, we get the expected mutual information, depicted as gray bars in the figure. As evident from the figure, there is a maximal mutual information given by physical limitations of crosstalk and noise. For large bin sizes with near-zero crosstalk between symbols, one can achieve very high mutual information of over $9$ bit per photon. Given the FWHM spot size of 8 pixels, we choose an $8\times8$\ binning achieving a value of $10.5$ bit of mutual information per detected photon.

To calculate the mutual information per {\it sent} photon, one has to take the losses in our setup into account. This includes the coupling losses into the single mode fibres of $55.1\%$, the diffraction losses at the SLM of $24\%$, the losses at the spectral filter of $30\%$ and the losses at the detector due to the limited quantum efficiency of $5\%$. This leads to a channel capacity of $0.1$ bit per photon.

\section{Discussion and Summary}
\label{Discussion}

\begin{figure}[htbp]
\centering
\includegraphics[width=7cm]{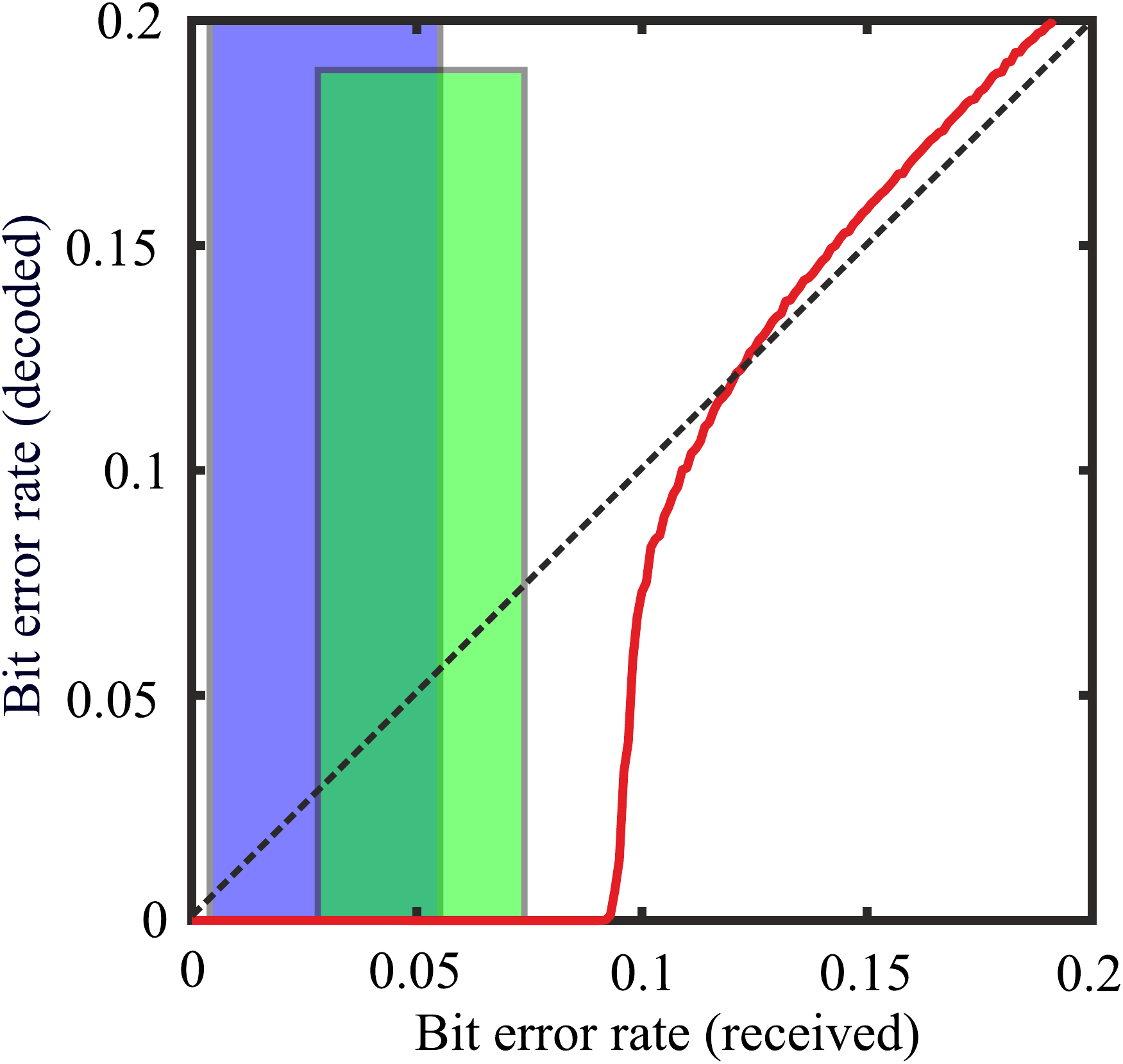}
\caption{The Bit Error Rate (BER) of the received bit string versus the BER of the bit string after performing error correction. The dashed diagonal line represents the result without error-correction. The vertical bars indicate the estimated BER of our experiment in case of $8\times8$ (green) and $12 \times 12$ (blue) binning. Their left and right edges indicate a signal-to-dark-count photon ratio of $100$ and $10$, respectively.}
\label{fig:LDPC}
\end{figure}

For useful communication, the errors within the received message have to be corrected. An efficient way of error correction is the Low-Density-Parity-Check-Code (LDPC) \cite{Gallager63low-densityparity-check}. In order to apply this error correction, all symbols of the set have to be translated into a bit string. Therefore we encode the $x$ and $y$ position of our symbols independently, each allocating half the bits. Since the dominant noise term is the crosstalk between the neighbouring symbols, we use a gray code \cite{gray1953pulse} for each direction. This causes only one bit flip for an erroneous detection by a neighbouring symbol either in $x$ or the $y$ direction.

Fig. \ref{fig:LDPC} shows the Bit Error Rate (BER) after error correction with LDPC versus the BER of the received bit string. The LDPC code was set to the half-rate LDPC used in digital television broadcast standard DVB-S.2. The coloured vertical bars indicate the estimated BER in case of $8\times8$ and $12 \times 12$ binning. The estimation takes the measured crosstalk between the symbols into account. Their left and right edges indicate a signal-to-dark-count photon ratio of $100$ and $10$, respectively. For the ICCD used in these measurements, this ratio is $10.07$. Other  commercially available ICCD have ratios close to $100$, explaining the choice of the other bound. Clearly, already standard error-correction code allows a practically error-free communication with the present system.

In conclusion, we demonstrate high-dimensional encoding of single photons reaching $10.5$ bit per photon. The capacity of this spatial encoding is only limited by the optics and the number of pixels on the detector. The channel capacity of $0.1$ bit can be increased by reducing the losses in the system. The main contribution to losses in our setup arises from the low quantum efficiency ($\sim 5\%$) of the ICCD, which can be improved to $\sim 30\%$ with different photocathode materials. This makes it possible to reach a signal-to-dark-count ratio of $100$ and would bring the measured values closer to its theoretical maximum. Our results are directly applicable to free-space line-of-sight communication. If the scrambling of wavefronts in multimode fibers can be controlled \cite{Cizmar:11}, a second and potentially more robust carrier for this high-dimensional encoding can be realised. A very promising direction for this work would then be the implementation of a large-spatial-alphabet encoding for quantum key distribution. 

\section*{Methods}
\subsection{Setup}
The setup is illustrated in Fig. \ref{fig:setup}. We use a Spontaneous Parametric Down-Conversion source \cite{PhysRevA.93.053817} to produce photon pairs. A \SI{790}{\nano\metre} mode-locked picosecond laser with a pulse repetition rate of \SI{76}{\mega\hertz} is frequency doubled to \SI{395}{\nano\metre} in a LBO crystal. The frequency-doubled light is focussed in a Periodically Poled Potassium Titanyl Phosphate (PPKTP) crystal, which spontaneously produces orthogonally polarized photon pairs at a wavelength of \SI{790}{\nano\metre}. The entangled photon pair is separated at a Polarizing Beam Splitter into two single-photon Fock states. One of these photons is sent to a Single Photon Counting Module acting as a herald, while the second photon is directed through a \SI{28.5}{\metre} single-mode fiber with a throughput of $47 \%$ to the encoding setup. For free-space communication, one needs an encoding device at the sender position and a decoding device on the receiver position. As encoding device we use a phase-only Spatial Light Modulator to modulate the wavefront of the single photons. By writing a blazed grating on the SLM, we change the reflection angle within an angular range of $0.8^{\circ}$ in vertical and horizontal direction with a diffraction efficiency of $76 \%$ in the first order. The Fourier transform of the SLM is imaged with a \SI{1}{\metre} focal length lens in 2f configuration onto a large-area spatially resolving photon-counting detector. A bandpass filter at $800$ $\pm$ \SI{40}{\nano\metre} is placed in front of the detector to block stray light. 

\subsection{Detector}
The decoding device should be able to measure the arrival of a single photon in a single shot on a large area. One technology that can potentially achieve this is the Intensified Charge-Coupled Device \cite{Jost1998,Abouraddy2001,Jedrkiewicz2004,Haderka2005,DiLorenzoPires2009,Aspden2013,Fickler2013}.
ICCDs provide nanosecond gating option, which reduces the amount of dark counts significantly and makes such an ICCD capable of heralded measurements, reducing dark counts to one per thousands of readouts. The dark counts of the ICCD have their origin in thermal electrons released by the photocathode of the ICCD. Additionally, residual gas atoms can be ionized by the electron avalanche within the Microchannel Plate (MCP) of the intensifier. These ions are accelerated towards the photocathode by the MCP bias voltage, releasing secondary electron avalanches. An ion impact causes many more electrons than an incoming photon does. This leads to a local signal increase on the ICCD camera which is brighter than the signal produced by a single photon. Therefore these spurious ion signals can be filtered out in postprocessing. Our model (Andor iStar A-DH334T-18u-A3) has a quantum efficiency of $5\%$. Each heralded photon from the SPDC source opens the ICCD gate for \SI{2}{\nano\second}. At each position corresponding to the center of one of the symbols, $6$ images with an exposure time of \SI{0.1}{\second} have been taken. Setting a herald rate of $\sim$ \SI{400}{\kilo\hertz}, we measured on average $7.3$ photon detections per position. Measuring the FWHM of the focus with a constant phase pattern on the SLM was found to be $7.9\pm 0.3$ pixels horizontally and $7.4\pm0.2$ pixels vertically.  

\subsection{Encoding}
The target position of the photon on the ICCD is determined by the horizontal and vertical diffraction angle of the grating on the SLM. Although a scan mirror could be used for spatial encoding, a SLM is a more flexible tool. Via holography, the phase and amplitude of a wavefront can be manipulated, allowing to use complex wavefronts. Moreover, the path of the light can be corrected for disturbances using wavefront-shaping methods \cite{Vellekoop:07}. The SLM is programmed with horizontal and vertical gratings to scan over the detection plane of the ICCD. In order to define the sent symbol, the position on the ICCD has to be mapped to separate symbols. For this reason, a grid is defined on the ICCD. The pixels of the ICCD are binned together in detection areas of $8\times 8$ pixels, forming an alphabet of $9072$ symbols. This rectangular map on the detection plane of the ICCD connects every detection area to an individual label,  numbered  from left to right and top to bottom.

\section*{Funding information}

European Research Council (ERC) (279248); Stichting voor Fundamenteel Onderzoek der Materie (FOM); Nederlandse Organisatie voor Wetenschappelijk Onderzoek (NWO).

\section*{Acknowledgments}

We would like to thank Klaus Boller, Boris \v{S}kori\'{c} and Willem Vos for support and discussions.

\bibliographystyle{apsrev4-1}

\end{document}